\documentclass[twocolumn,showpacs,superscriptaddress,prc]{revtex4-1}
\usepackage{dcolumn,pstricks,amssymb}
\usepackage[intlimits]{amsmath}
\ifx\pdfoutput\undefined
\usepackage[dvips]{graphicx}
\else
\usepackage[pdftex]{graphicx}
\usepackage{epstopdf}
\fi
\usepackage{bm}
\usepackage{ulem}
\usepackage{lipsum}
\usepackage{graphicx}

\newcommand\Tstrut{\rule[2ex]{0pt}{2pt}}         
   
\begin{document}
\title{\boldmath Precision Beta Decay as a Probe of New Physics}
\author{Vincenzo Cirigliano}
\affiliation{
Theoretical Division, 
Los Alamos National Laboratory,
Los Alamos, NM 87545, USA}
\author{Alejandro Garcia} 
\affiliation{Department of Physics and Center for Experimental Nuclear Physics and Astrophysics, University of Washington,Seattle, Washington 98195, USA}
\author{Doron Gazit} 
\affiliation{Racah Institute of Physics, The Hebrew University of Jerusalem, 9190401 Jerusalem, Israel}
\author{Oscar Naviliat-Cuncic}
\affiliation{National Superconducting Cyclotron Laboratory and Department of Physics and Astronomy, Michigan State University, East Lansing, MI 48824, USA}
\author{Guy Savard} 
\affiliation{Physics Division, Argonne National Laboratory, Lemont, Illinois 60439, USA
Department of Physics, University of Chicago, Chicago, Illinois 60637, USA}
\author{Albert Young}
\affiliation{Department of Physics, North Carolina State University, Raleigh, North Carolina 27695, USA, and Triangle Universities Nuclear Laboratory, Durham, North Carolina 27708, USA}
\maketitle
\begin{widetext}
\section{Overview and Summary}
Theoretical and experimental advances in recent years have paved the way for a significant step in our understanding of the charged-current weak interaction in the nucleus and opened opportunities to push the sensitivity of a select group of high precision probes for new physics beyond the reach of the LHC. These advances have come with the refinement of experimental techniques, such as magnetic trapping of ultracold neutrons, trapping of radioactive ions and atoms, and new methods for beta spectroscopy, and also through a change in the theoretical landscape, with the introduction of a model independent, effective field theoretical analysis of beta decays and LHC constraints, high precision lattice calculations of the nucleon form factors, the advent of systematic methods for evaluating nuclear structure and operators, and most recently, with a new analysis of the electro-weak radiative corrections. This paper summarizes discussions on workshops at Amherst, in November of 2018, and at Trento, in April of 2019. The primary goal of the workshops was to evaluate beta-decay experiments for their sensitivity to beyond standard model physics, identifying critical theory needs and the resources to address the problems.
This document covers the experimental approaches and theoretical inputs required to achieve the highest sensitivity to new physics, taking into account both 
existing data and future sensitivity estimates from the LHC.

The standard model (SM) description for beta decay involves couplings with vector (V) and axial-vector (A) Lorentz structure 
with a pure left-handed, V-A form. The review conclusions can be broken down into the analysis of beyond standard model (BSM) couplings with $V\pm A$ structure
and those with exotic scalar (S) or tensor (T) structure. 

Through model-independent, effective field theory analysis, the CKM matrix unitarity test is firmly established as one of our most sensitive and useful probes for Beyond Standard Model (BSM) physics, with experimental uncertainties for the superallowed, $0^+\rightarrow 0^+$ $ft$ values providing $<10^{-3}$ sensitivity to new couplings and with radiative corrections providing the largest uncertainty in the unitarity test until the past year. 
Recent advances in radiative-correction calculations
have led to a different landscape, with a factor of two smaller errors due to these effects and a roughly 4$\sigma$ discrepancy of the unitarity  sum with the SM. 
This new landscape strongly motivates focused efforts on beta-decay research. Given that the unitarity sum is one of our most precise probes for new physics, resolution of this issue is critical in establishing the validity of our current understanding of the standard model. 
The most promising paths to progress 
requires
both experimental and theoretical efforts. On the experimental side, data from the neutron at the 0.02\% precision level is a high priority. From the theory side, the new approach to radiative corrections must be confirmed and vetted. In addition, a multi-pronged effort is needed to provide a rigorous analysis of nuclear-structure effects and to integrate many-body effects into radiative correction calculations.

In terms of searching for exotic couplings (i.e couplings with S and T 
structure), high-precision measurements of beta spectra and of some correlation parameters in light nuclei and the neutron were identified as most promising to reach sensitivity comparable or beyond the LHC. Experiments have goals of searching for the Fierz interference term with uncertainties $b < 10^{-3}$ implying sensitivity to new physics scales beyond 10 TeV. 
These experiments will only reach their ultimate sensitivity provided sufficient progress in theory (for the nuclear decays) is achieved to calculate the SM contributions, including recoil order effects and radiative corrections. The outlook for nuclear-structure theory was considered in further depth at the ECT (Trento) workshop. The current status of ab-initio calculations suggests that not only are the required calculations feasible, but can be approached in a self-consistent framework with a well-defined expectation for the accuracy of the structure model. Given the scope of these projects, the structure problems therefore do seem solvable, but not without investment to support young talent in the field.

The experimental data for the neutron lifetime will be addressed in the coming years. In addition to the relevance for the CKM matrix mentioned above, this parameter plays an important role in our understanding of big bang nucleosynthesis, solar fusion and neutrino cross-sections and searches for S and T currents. The past two years have seen experimental results reinforcing concerns that measurements made of the decay protons in cold neutron beams differ from the lifetime extracted from stored ultracold neutrons by $8.3 \pm 2.0$~s, a large enough discrepancy to generate an international effort to determine the source of this difference.   

This document does not address searches for Time Reversal Symmetry Violation or measurements of the electron neutrino mass. The structure of this document is to first review the basis for our sensitivity estimates for new physics in Sec.~\ref{sec: hep motivation}, deal specifically with the tests of unitarity of the CKM matrix in Sec.~\ref{sec: ckm matrix}, and then very briefly review, in Sec.~\ref{sec: experiments}, the status and plans for high precision beta decay experiments.  This sets the stage to identify the critical inputs required in this experimental program to achieve their targeted sensitivity in Sec.~\ref{sec: theory-req}, the current status and strategies of nuclear theory in Sec.~\ref{sec: nuclear theory next decade} and provide our conclusions in Sec.~\ref{sec: conclusions}.

Overall, we see the next decade as a real opportunity to push a broadband 
sensitivity to new physics for the full set of V, A, S and T 
couplings to energy scales at the level of, or higher than those accessible by the LHC. The envisioned research program also has the necessary ingredients to either resolve or sharpen the current CKM unitarity and neutron lifetime issues.  
\end{widetext}

\section{The landscape: energy versus precision frontiers}

\label{sec: hep motivation}
The Standard Model of particle physics (SM) synthesizes our knowledge of the workings of nature at its most fundamental level. Despite its huge success over the last 40 years, the SM has a number of empirical shortcomings (origin of baryon asymmetry, nature of dark matter, neutrino masses, to name a few), and leaves several theoretical  
questions unanswered (such as what is the 
origin of three generations and their mixing pattern,  
is there a unifying principle beyond the observed interactions, 
what stabilizes the hierarchy between weak and Plank scales, 
etc.).
Measurements at the LHC represent the {\it energy frontier} and 
provide a direct
avenue towards finding hints of new physics at a high energy scale~\cite{peskin:16}. Apart from the spectacular discovery of the Higgs, no other signs of new physics have shown up so far at the LHC, so the questions mentioned before remain unanswered. 

The sensitivity to new physics in beta decay comes at the {\it precision frontier}, via tests of the unitarity of the CKM matrix, and by precision measurements of beta spectra and correlations sensitive to the chiral structure of the weak interactions (see \cite{go:18} and references therein). The impact of precision beta decay measurements can be assessed in the context of effective field theory~\cite{Cirigliano:2009wk,
Bhattacharya:12,Cirigliano2013,ci:13,Alioli:2017}. The most sensitive observables are those leading to interference of the SM currents with the new physics. Thus, the new physics can show up as additional gauge-fermion vertices ($W$-$f$-$f^\prime$) or new contact four-fermion interactions generated by exchange of heavy particles, either at tree or loop level. 
The resulting BSM interactions can be parameterized 
in terms of coupling strengths $G_F \times \epsilon_\alpha$, 
where $\epsilon_\alpha$ are dimensionless parameters and 
$\alpha \in \{L,R,S,P,T \}$ labels the Lorentz structure of the interaction. The BSM physics
can generate additional contributions to the vector and axial vector couplings of the standard model, with coupling strengths
coupling constants $\epsilon_L$ and $\epsilon_R$, 
as well as chiralty-flipping currents --scalar, pseudo-scalar, and tensor, with couplings $\epsilon_S$, $\epsilon_P$, and $\epsilon_T$, respectively). 
The dimensionless couplings are related to the new physics scale $\Lambda$ via 
\begin{eqnarray}
\epsilon_i \approx \left( \frac{v}{\Lambda} \right) ^2
\end{eqnarray}
with $v \approx 174$~GeV, the vacuum expectation value. Thus, experiments reaching sensitivity $\epsilon_i <10^{-3}$ can explore the $\Lambda = 5-10$~TeV range, 
not fully explored at the energy frontier,
where many models for new physics provide observable 
effects (see~\cite{Bauman:2012fx,go:18} and references therein). 
\vspace{0.5cm}

\section{Tests of unitarity of the CKM matrix}
\label{sec: ckm matrix}
The check for the unitarity of the CKM matrix is an example that has already reached experimental precision to test for $\epsilon_{L,R}$ at an interesting level and has recently presented surprises. This observable has competition only from precision electroweak data, for which no improvement is expected until the next linear collider is built, but not from the  LHC~\cite{Alioli:2017}.

In the SM unitarity test $|V_{ud}|^2 + |V_{us}|^2 + |V_{ub}|^2 = 1$ the largest contribution comes by far from $V_{ud}$, while the error budget receives comparable contributions from both $V_{ud}$ and $V_{us}$. Therefore,
nuclear data has played a central role in this 
high-precision test of the SM. Presently $V_{ud}$ is extracted with the highest precision from $0^+ \rightarrow 0^+$ decays from nuclei ranging from $^{10}{\rm C}$ to $^{74}{\rm Rb}$~\cite{Hardy:2014qxa}.

This data set also provides a stringent test of CVC and a limit on $\epsilon_S$. Using data from nuclear decay, however, imposes a demand on theory to evaluate radiative and isospin-symmetry breaking corrections with unprecedented precision and taking uncertainties in nuclear theory into consideration. The nuclear structure dependence of these calculations have been validated to a degree by the consistency of the results obtained on the various isotopes in the data set but missing overall factors could still be present.

Recently, a new way of evaluating a part of the radiative correction that was thought to be critical in limiting the precision of the test, the so-called $\gamma-W$-box correction for the nucleon, has been performed~\cite{seng-prl:18}. With respect to the previous calculation~\cite{PhysRevD.70.093006}, it has yielded both a reduction in uncertainty of approximately a factor of two as well as a different central value. The shift is significant and an independent confirmation would be desirable
but, taken at face value, and using the latest recommendation for $V_{us}$~\cite{flag:19}, would lead to a $>4 \sigma$ discrepancy with unitarity. 
Ref.~\cite{seng:18} shows that a consistent and precise evaluation of the $\gamma-W$ box for nuclei is non-trivial. They make an interesting connection to inclusive neutrino scattering, pointing out that better data would allow for a more accurate estimation. In addition, a path for a lattice calculation has recently been proposed~\cite{PhysRevLett.122.211802}.
Meanwhile Marciano has reported at workshops calculations yielding values roughly in-between the previous~\cite{PhysRevD.70.093006} and the results of Seng et al.~\cite{seng-prl:18}.
Refs.~\cite{seng:18,Gorch:19} explore the application of the dispersive method to evaluating the $\gamma W$-box on nuclei and accounting for the lower part of the nuclear excitation spectrum. 
Ref.~\cite{seng:18} addresses the correction that is uniform throughout the decay electron spectrum, while Ref.~\cite{Gorch:19} questions the validity of the usual approximation that assumes that nuclear structure does not alter the shape of the decay spectrum with an explicit calculation. These studies are not fully conclusive since both references operate with a free Fermi gas model calculation of only one part of the nuclear excitation spectrum, the quasi-elastic contribution. It is found that the modifications of the energy-independent and energy-dependent corrections nearly cancel each other 
in the ${\cal F}t$ values. Nevertheless, the individual shifts are large, about double the size of the commonly accepted uncertainty of the analysis of Hardy and Towner.

Discussions during recent workshops at Amherst~\footnote{Workshop on ``Beta Decay as a Probe of New Physics'' at Amherst Center for Fundamental Interactions, Nov. 1-3, 2018; https://www.physics.umass.edu/acfi/seminars-and-workshops/beta-decay-as-a-probe-of-new-physics}, and at Trento~\footnote{Workshop on ``Precise Beta Decay Calculations for Searches for New Physics'' at European Centre for Theoretical Studies in Nuclear Physics and Related Areas (Trento, Italy), Apr. 8-12, 2019; https://indico.ectstar.eu/event/42/timetable/\#20190408} show optimism for calculating these effects with the needed accuracy in light nuclei, such as $^{10}{\rm C}$ and $^{14}{\rm O}$, where ab-initio calculations (see, e.g.~\cite{navratil:97,Pastore:2017uwc}) can be performed. Alternatively, the additional terms shown in Ref.~\cite{Gorch:19} introduce a distortion of the beta spectrum and a $Q$-value dependent shift in the overall decay rate that could be observed with an improved nuclear data set. More fundamentally, these new calculations show that it is possible to extract the nucleon and nuclear structure dependent corrections within the same framework, removing potential issues with the separation that was introduced in previous methods.

$V_{ud}$ can also be extracted from neutron beta decay. In this case,  the nuclear structure dependent corrections disappear and theoretical uncertainties are limited only by that on the newly recalculated ``nucleon $\gamma-W$-box'' correction, which it shares with the nuclear decays.
The uncertainty on this new calculation has been reduced and in the case of the neutron the dominating uncertainties are experimental, mainly limited by the determination of the neutron lifetime and other experiments aimed at determining the axial nucleon form factor, $g_A$. A world-wide program to address apparent differences observed between measurements of the neutron lifetime as determined from experiments using beams of cold neutrons versus those using ultra-cold neutrons (UCNs) has been established. Sec.~\ref{sec: experiments} presents prospects on expected progress in the neutron beta-decay lifetime front, as well as in experiments searching for breaking of the chiral structure of the SM. 

Ideally, an accurate determination of $V_{ud}$ from both the nuclear decays and the neutron (and if possible the pion) would further improve sensitivity to BSM physics~\cite{Bhattacharya:12,go:18}.

\section{Experimental outlook: improvements in the next decade}
\label{sec: experiments}
A large part of the effort in high precision beta decay measurement has focused on decay rates and angular correlation measurements. One of the key drivers has been the potential impact these measurements have on tests of the CKM matrix unitarity. Presently, high precision results are derived from superallowed decays. As mentioned in Sec.~\ref{sec: ckm matrix}, progress on the superallowed decays in the next decade will likely be driven by theoretical developments. 
Neutron beta decay measurements are poised to achieve sensitivity competing with the superallowed decays over the next decade. Measurements of $ft$-values from nuclear mirror decays could also contribute allowing more probes of the potential sources of the current CKM discrepancy, as well as constraining BSM physics scenarios.

An emphasis is being placed on extracting high-precision beta energy dependence of the angular correlations, and new measurements of beta spectra. 
Below we briefly describe planned angular correlation measurements, spectral measurements, and the neutron lifetime.

\subsection{Precision experiments of correlations in nuclear and neutron decays}
The most precise measurements in neutron decay now have precisions comparable to the potential discrepancy which has emerged for CKM unitarity. Furthermore, the next generation of experiments (presented in Tables~\ref{tab:corr_and_spect} and ~\ref{tab:lifetime}) can weigh in at the precision levels of the new theoretical uncertainties for the EW-correction. As these nuclear-structure independent constraints sharpen, a real opportunity will be available to triangulate on the potential origins of a unitarity violation, should it persist. Specifically, two angular correlation experiments (Nab\cite{Pocanic2009} and one using PERC\cite{Dubbers:2007st}) are planned with precision levels for the axial coupling constant at or below the 0.025\% level, and at least seven neutron lifetime experiments are planned with nominal sensitivities for the lifetime at the 0.3~s or lower, which should result in a precision for $|V_{ud}|^2$ below roughly the 0.04\% during the next decade.

Neutron angular decay experiments provide unique sensitivity to the axial coupling constant, $g_A$. The data set for the axial coupling constant appears to be evolving, with most recent experiments in agreement, but with a larger magnitude for $g_A$ than those before PERKEO II.  In the short term, two experiments should report values of $g_A$ through measurement of the $\beta -\nu$ correlation (aCORN\cite{Darius:2017a} and aSPECT\cite{Schmidt:2019}) at a relevant precision level (roughly 1\%) with the potential to clarify the situation.  These experiments  have very different systematic error budgets from the $\beta$-asymmetry measurements which dominate the data set for $g_A$, and therefore may have significant impact. This axial coupling constant can now be directly compared with standard model predictions from the lattice, providing another test for modifications of the standard model expectations for axial currents, significantly more sensitive than those one can produce from LHC.~\cite{Alioli:2017}.  The precision of this test is currently driven by theory, with the quoted precision of lattice calculations of $g_A$ from 1\%\cite{gA-nature:2018} to 4\%\cite{Gupta:2018a}, with some controversy as to what precision level is realistic. On the other hand, optimistic theoretical predictions suggest a factor of roughly 2 to 4 is immediately available in this parameter, as lattice groups gain access to state-of-the-art super-computing resources. These constraints are new and very well motivated, as beta decay provides the definitive value for the $g_A$ and measurements at the LHC are not competitive for BSM extensions with A symmetry~\cite{Alioli:2017}.

Mirror nuclei also provide an independent extraction of $V_{ud}$\cite{PhysRevLett.102.142302}, with $^{19}$Ne and $^{37}$K now providing $|V_{ud}|$ at about the 0.2\% level~\cite{PhysRevLett.120.062502} with recent progress on essentially all key components of the data for these systems, where historically the precision of the angular correlation measurements has dominated the uncertainty budgets. As the focus on theoretical analysis of the nuclear systems sharpens, low $Z$ mirror systems such as $^{13}$N, $^{17}$F and $^{19}$Ne may provide important cases for systematically checking nuclear structure corrections.  Major experimental programs are envisioned at TAMU and Notre Dame to develop high precision mirror decay angular correlation measurements of the beta-neutrino correlation.

Taken as a whole, these high precision measurements can be analyzed to constrain generic standard model extensions (the whole range of SM extensions with V, A, S and T-symmetries). In this context, the mirror decays and the neutron play complementary roles, with the superallowed decays providing sensitivity to standard model extensions with scalar and vector symmetries, and mirror decays and the neutron sensitive to (depending on the particular decay) all possible Lorentz symmetry extensions. For example, considering the decay rates and angular correlation measurements of the neutron, superallowed and Gamow-Teller decays\cite{Gonzalez-Alonso:2019}, one derives generic limits at roughly a factor of two higher energy scale than the LHC for BSM extensions with vector and axial vector symmetry, comparable to the LHC for BSM scalar interactions and somewhat lower than the LHC for tensor interactions.  Another independent analysis of a broader set of nuclear data (including the mirror decays) suggests tensor limits comparable to those from the LHC.  The next generation of measurements can certainly improve these limits.  In particular, planned correlation measurements on superallowed nuclei should help produce independent constraints from the decay rate data (used for unitarity analysis), strengthening the unitarity probe.  For tensor interactions, the energy scale should be pushed to above 11 TeV with the planned program of measurements and with even stronger limits possible if complementary positron mirror decays are also characterized at the 0.02\% level.

It is worth noting that the BRAND experiment\cite{Bodek:2019}, currently in an R\&D phase, plans to determine a collection of 11 angular correlations in neutron decay, involving measurements of the electron and proton momenta, and the electron and neutron spins, with sensitivities at roughly the $10^{-3}$ level or below for all of these observables. Sensitivity at this level is planned for the European Spallation Source (ESS), for which a particle physics beam line is proposed, and which could be available as early as 2025. 

Tables~\ref{tab:nuclear decays} and~\ref{tab:corr_and_spect}, shown below, summarize expectations for progress in measurements of $\beta$-decay correlations over the next decade. 
\begin{table*}
\caption{List of nuclear $\beta$-decay correlation experiments in search for non-SM physics
\footnote{Experiments specifically searching for time-reversal symmetry violation not listed here}}
\begin{ruledtabular}
\begin{tabular}{lllll}
 Measurement & Transition Type & Nucleus & Institution/Collaboration &Goal \\ 
\hline
$\beta-\nu$   &F                  & $^{32}{\rm Ar}$                   &   Isolde-CERN  &  $0.1$ \%  \Tstrut \\
$\beta-\nu$   &F                  & $^{38}{\rm K}$                    &   TRINAT-TRIUMF &  $0.1$ \% \\
$\beta-\nu$   &GT, Mixed  &  $^{6}{\rm He}$, $^{23}{\rm Ne}$              &   SARAF &  $0.1$ \% \\
$\beta-\nu$   &GT                & $^8{\rm B}$, $^8{\rm Li}$ &   ANL              &  $0.1$ \% \\
$\beta-\nu$   &F           & $^{20}{\rm Mg}$, $^{24}{\rm Si}$, $^{28}{\rm S}$, $^{32}{\rm Ar},...$          &   TAMUTRAP-Texas~A\&M &  $0.1$ \%  \\
$\beta-\nu$   &Mixed           & $^{11}{\rm C}$, $^{13}{\rm N}$, $^{15}{\rm O}$, $^{17}{\rm F}$                    &   Notre Dame &  $0.5$ \% \\
$\beta$ \& recoil & Mixed                  & $^{37}{\rm K}$                    &   TRINAT-TRIUMF &  $0.1$ \% \\
asymmetry   &                   &                    &    &   \\
\end{tabular}
\label{tab:nuclear decays}
\end{ruledtabular}
\end{table*}
\begin{table*}
\caption{Summary of planned neutron correlation and beta spectroscopy experiments}
\begin{centering}
	\begin{tabular}{ l r r r r r r r}
		\hline\hline \multicolumn{1}{c}{Measurable} &\multicolumn{1}{c}{Experiment} & \multicolumn{1}{c}{Lab} & \multicolumn{1}{c}{Method} & \multicolumn{1}{c}{Status} & \multicolumn{1}{c}{Sensitivity} &\multicolumn{1}{c}{Target Date} \\
                &&&&& (projected) &  \\ \hline\hline
                $\beta-\nu$ & aCORN\cite{Darius:2017a} & NIST & electron-proton coinc. & running complete & ~1\% & N/A \\
                $\beta-\nu$ & aSPECT\cite{Schmidt:2019} & ILL & proton spectra & running complete & $0.88$\% & N/A \\
                $\beta-\nu$ & Nab\cite{Pocanic2009} & SNS & proton TOF & construction & $0.12$\% & 2022 \\
                $\beta$ asymmetry & PERC\cite{Dubbers:2007st} & FRMII & beta detection & construction & $0.05$\% & commissioning 2020 \\
                11 correlations & BRAND\cite{Bodek:2019} & ILL/ESS & various & R\&D & $~0.1$\% & commissioning 2025 \\
                $b$ & Nab\cite{Pocanic2009} & SNS & Si detectors & construction & $0.3$\% & 2022 \\
                $b$ & NOMOS\cite{Konrad:2018_privcomm} & FRM II & $\beta$ magnetic spectr. & construction & $0.1$\% & 2020 \\
                \hline \hline\\
         \end{tabular}
\end{centering}
\label{tab:corr_and_spect}
\vspace{-0.05in}
\end{table*}


\subsection{Spectral measurements}
Precise measurements of beta spectra are very sensitive probes of non-$V-A$ currents. If scalar or tensor interactions existed, they would generate a distortion of beta spectra, called the Fierz interference, proportional to $m/E$. Measurements with precision $b < 10^{-3}$ for GT transitions yield sensitivity to T currents:
\begin{eqnarray}
\epsilon_T &\lesssim& 1.5 \times 10^{-4} 
~~{\rm or}\nonumber \\
\Lambda &\gtrsim& 14~{\rm TeV}.\nonumber
\end{eqnarray}
For this reason there is renewed interest in spectroscopic methods. 

New experiments to measure the beta spectra will exploit well developed spectroscopic methods such as ``hermetic'' Si detector systems (MiniBETA, the LANL $^{45}{\rm Ca}$ experiment and Nab) as well as magnetic spectrometers (NOMOS). Other experiments will further improve the calorimetric technique with implanted radioactive beams developed at NSCL\cite{nav16,huy18}, and there are ongoing efforts to use cyclotron radiation emission spectroscopy\cite{p8:prl} to develop new handles on the sources of systematic error in spectroscopy, and offer the hope of pushing below the $10^{-3}$ level in Fierz terms.

Table~\ref{tab:spectral} shows a list of experiments aiming at beta-spectra measurements with nuclear systems.  The neutron decay beta-spectra measurements are incorporated in Table~\ref{tab:corr_and_spect}.  

For neutron decay, there are no expected theoretical uncertainties above the $10^{-4}$ level, strongly motivating neutron decay measurements, but these advantages are balanced by the neutron being rather insensitive to scalar interactions and the difficulties of matching the availability decay rates of some equally sensitive nuclear decays such as $^{6}$He.  A recent overview of capabilities of standard approaches to predict beta spectra\cite{Hayen:2018sd} indicates relative uncertainties at the level of a few $\times 10^{-4}$. Searches for chirality-flipping interactions aiming at sensitivities beyond $10^{-3}$ will need improvements in calculations. This should be feasible, particularly for lighter nuclei, where ab-initio calculations can reach the needed precision.
\begin{table*}
\caption{List of nuclear $\beta$-decay spectral measurements in search for non-SM physics
\footnote{Experiments specifically searching for time-reversal symmetry violation not listed here}}
\begin{ruledtabular}
\begin{tabular}{lllll}
 Measurement & Transition Type & Nucleus & Institution/Collaboration &Goal \\ 
\hline
$\beta$ spectrum   & GT & $^{114}{\rm In}$ &  MiniBETA-Krakow-Leuven   &  $0.1$ \%  \Tstrut\\
$\beta$ spectrum   &GT & $^6{\rm He}$ &  LPC-Caen   &  $0.1$ \%  \Tstrut\\
$\beta$ spectrum   &GT       & $^6{\rm He}$, $^{20}{\rm F}$ &  NSCL-MSU        &  $0.1$ \%  \Tstrut\\
$\beta$ spectrum   &GT, F, Mixed & $^6{\rm He}$, $^{14}{\rm O}$, $^{19}{\rm Ne}$ &  He6-CRES   &  $0.1$ \%  \Tstrut\\
\end{tabular}
\label{tab:spectral}
\end{ruledtabular}
\end{table*}

\subsection{Neutron decay lifetime}
\label{sec:neutron lifetime}
As described in Sec.~\ref{sec: ckm matrix}, a central issue for a precise extraction of $V_{ud}$ from neutron beta decay is the experimental status of the neutron lifetime\cite{RevModPhys.83.1173,Mumm605}. This quantity also plays a role in high precision predictions of Big Bang nucleosynthesis, solar fusion rates and neutrino cross-sections.  
The global lifetime data-set is dominated by measurements of ultracold neutrons (UCN) stored in material and magnetic traps, with the most precise of material trap experiment (gravitrap) reporting values of $878.5 \pm 0.7 ({\rm stat}) \pm 0.3 ({\rm sys})$~s\cite{PhysRevC.78.035505} at ILL and the most precise magnetic trap experiment (UCN$\tau$) reporting $877.7 \pm 0.7 ({\rm stat}) +0.4/–0.2 ({\rm sys})$~s\cite{Pattie:2018} at LANL. The average of recent UCN measurements is 879.5(7)~s, with the uncertainty expanded to account for scatter, in sharp contrast with a well-established program of cold neutron beam measurements performed at NIST\cite{PhysRevLett.111.222501}.   These cold neutron beam measurements determine the absolute neutron beta decay rate by counting decay protons in a variable volume Penning trap and measurements of the neutron density, with a neutron lifetime (averaged over two similar experiments) of 887.8(2.0)~s. This discrepancy has already spurred significant investments over the next decade, involving an ongoing program at NIST with planned sensitivity below 2~s using the existing experimental apparatus (BL2) and a major upgrade planned to begin commissioning in 2023 (BL3).  In parallel, the UCN$\tau$ experiment is also developing a concrete plan for staged upgrades of the existing apparatus, with current runs targeting uncertainties around 0.25~s evolving ultimately to an experiment optimally matched to the LANSCE UCN source production and a factor of 4 improvement in the statistical uncertainty. The gravitrap experiment has a goal of below 0.3~s for its current efforts as well.

In addition to these leading experiments, there is a very large community of physicists developing new measurements.  These experiments include a cold neutron beam experiment, targeting 1~s precision, which measures the neutron density and beta decay rate in a time-projection chamber (JPARC-TPC), with an upgrade planned for the future to implement an ``entraining'' axial magnetic field for the charged particles produced in the TPC (LINA) which is targeting 1~s precision at present.  They also include four magnetic trap experiments which explore different loading, population measurement and spectral conditioning methods.  For example, the group of Ezhov was the first to report neutron lifetime measurement with magnetically trapped ultracold neutrons\cite{ezhov05,Ezhov:2018a}, and is currently constructing a new permanent magnet trap with a goal of 0.3~s precision\cite{Ezhov:2016a}.  The Ezhov trap, TauSpect\cite{Ries_aSPECT_lifetime}, PENeLOPe\cite{Picker:2019} and HOPE\cite{LEUNG2009181} should significantly broaden our assessment of the systematic errors associated with the magnetic trapping technique. Overall, the community goal for the next decade is to establish a consistent value for the beta decay lifetime of the neutron at the 0.3~s or below, with the most precise measurements aiming for 0.1~s level precision.  We note that an R\&D project (PROBE) is also on-going to develop new absolute beta decay rate experiments using UCN, as a cross-check of cold neutron proton counting experiments\cite{Tang:2019}.  
\begin{table*}
\begin{centering}
\caption{Summary of planned neutron lifetime experiments}
	\begin{tabular}{ l l r r r r r}
		\hline\hline \multicolumn{1}{c}{Experiment} & \multicolumn{1}{c}{Lab} & \multicolumn{1}{c}{Method} & \multicolumn{1}{c}{Status} & \multicolumn{1}{c}{Sensitivity} & \multicolumn{1}{c}{Sensitivity} &\multicolumn{1}{c}{Target Date} \\
                &&&& (current) & (projected) & (projected) \\ \hline\hline
\vspace{0.05in}
                Gravitrap\cite{Serebrov:2018a} & ILL &UCN Material Trap & running & $0.92$~s & $0.3$~s & N/A \\
                UCN$\tau$\cite{Pattie:2018} & LANL & UCN Magnetic Trap & running & $0.7$~s & $<0.3$~s & 2020 \\
                HOPE\cite{LEUNG2009181} & ILL & Hybrid Trap & running & $15$~s & $0.3$~s & N/A \\
                TauSPECT\cite{Ries_aSPECT_lifetime} & Mainz & UCN Magnetic Trap & construction && $0.3$~s & commissioning 2019 \\
                PENeLOPE\cite{Picker:2019} & N/A & UCN Magnetic Trap & construction && $0.1$~s & commissioning 2020  \\
                Ezhov Trap\cite{Ezhov:2016a} & ILL & UCN Magnetic Trap & construction && $0.3$~s & N/A \\
                BL2\cite{Hoogerheid:2019} & NIST & CN beam proton det & running && below $2$~s &  running till 2021 \\
                BL3\cite{Wietfeldt:2019} & NIST & CN beam proton det & funded && $0.3$~s & commissioning 2023 \\

		        J-PARC TPC\cite{Nagakura:2017a}& J-PARC & CN beam TPC & running & $10$~s & $1$~s & N/A \\
                PROBE\cite{Tang:2018a} & LANL & branching ratio & R\&D && $1$~s & 2021 \\ 
                \hline \hline\\
         \end{tabular}
\end{centering}
\label{tab:lifetime}
\end{table*}

\section{Requirements from Nuclear Theory}
\label{sec: theory-req}
One source of uncertainty that has bearings in comparisons between LHC and nuclear data are nucleon charges, namely, form factors evaluated at zero momentum transfer. There has been recently important progress in lattice computation of the scalar, tensor, and axial charges, $g_S$, $g_T$, and $g_A$ (see, for example, Ref.~\cite{Green2012,PhysRevD.95.114514,Gupta:2018a,gA-nature:2018} and references therein). The uncertainty on $g_T$ is at the 5\% level and sufficient for the present needs, but the uncertainty on the scalar is approx. 10\% ~\cite{Gupta:2018a}.
A further reduction of about 2 should be attainable and will be welcome. 
The uncertainty on $g_A$ is subject of debate in the literature. 
While it is agreed that the uncertainties are lower than $3\%$, only one group reported an uncertainty  at the level of 1\% \cite{Chang:2018bw}, that others  consider to not account for all systematic uncertainties~\cite{Gupta:2018a}.
We look forward to a resolution of the discrepancies and improvements on the uncertainties. 
This would allow for more sensitivity to identify new physics 
with $V+A$ chiral structure ($\epsilon_R$), 
through a comparison of the calculated and measured 
values of $g_A$.

Regarding tests of the unitarity of the CKM martix, Hardy and Towner~\cite{Hardy:2014qxa}
carried out careful analyses of data and calculations of corrections needed to extract $V_{ud}$ from superallowed Fermi transitions with precision improving over many years of work. It is remarkable work that rendered a sensitive probe for new physics. The approach followed by Hardy and Towner\cite{Hardy:2014qxa} assumes a particular decomposition of radiative and isospin-breaking corrections, with a logic that makes much sense in the context of a traditional phenomenological description of the complicated nuclear-structure problem. Radiative and isospin-breaking corrections are calculated with the help of shell-model calculations and Saxon-Woods approximations to estimate the radial overlap of initial and final wave functions, and applied to nuclei from $^{10}{\rm C}$ to $^{74}{\rm Rb}$. The uncertainties are estimated {\it a posteriori} by gauging the constancy (i.e. independence on $Z$) of the corrected ${\cal F}t$ values. This approach was developed with patience to critically assess the experimental data and clever ideas to find ways of simplifying and refining the nuclear calculations. The results have shown a remarkable level of consistency, so far not reached by any alternative method\cite{PhysRevC.82.065501}. The method, however, makes significant approximations, such as treating part of the radiative ``nuclear $\gamma-W$-box contribution'' as dominated by Gamow-Teller transitions followed by M1 decays feeding the final state, with associated form factors that are extracted from comparisons to single-transitions data. The method does not allow for estimation of uncertainties in an ab-initio way. An important theory need is thus the solution to the problem of radiative and isospin-breaking corrections performed within an ab-initio model and with an EFT approach that would allow one to solve the problem consistently and, if possible, accounting for uncertainties in a realistic way. More details on the prospects are given in Sec.~\ref{sec: nuclear theory next decade}.

In searches for contributions of exotic chirality-flipping interactions, the accurate description of SM observables requires also the proper inclusion of small effects, such as radiative corrections and contributions of induced hadronic form factors, like weak magnetism or the induced tensor.
In contrast to the extraction of $V_{ud}$ from superallowed transitions, the searches for exotic interactions through measurement of correlation parameters or through spectrum shape measurements do not require knowing absolute matrix elements or corrections which do not affect the dependence on the $\beta$-particle energy.

For example, the effect of radiative corrections and of the induced tensor form factor in measurements of $\beta-\nu$ angular correlations in $^6$He and $^{32}$Ar, has been discussed in Ref.~\cite{Glu98}. The inclusion of these effects requires often Monte-Carlo simulations of the experimental conditions as well as the consideration of the particular decay properties of the selected nuclei.

Another example is the impact of such corrections in measurements of the shape of $\beta$ energy spectra. The regular order-$\alpha$ radiative correction generally assumes that photons from inner-bremsstrahlung are fully distinguishable from $\beta$ particles and are not detected. Whereas such conditions generally apply to magnetic spectrometers and gas detectors, and will likely apply to the new CRES technique, the assumption is not valid for calorimetric techniques, which require dedicated Monte-Carlo simulations.

Moreover, in well selected transitions, the largest hadronic form factors for the analysis of the shape of $\beta$ energy spectra are the weak magnetism, $b_{WM}$, and the induced tensor, $d^I$. The selection of a particular nucleus is then crucial for limiting the impact of uncertainties in these form factors. Interesting candidates are $\beta$ decay transitions in isospin triplets, because the weak magnetism form factor can then be related to the width of the isovector part of the analogous M1 electromagnetic transition through the strong form of CVC, and can be calculated from other experimental data. In the notation of Ref.~\cite{Holstein:1974gl}, the current value of the weak magnetism form factor calculated from CVC for $^6$He is $b_{WM} = 68.22(79)$, where the uncertainty is dominated by the width of the M1 transition in $^6$Li. The sensitivities of the Fierz term, $b_{GT}$, to the form factors in this transition, deduced from Monte-Carlo simulations, are listed in Table.~\ref{tab:Deltas_bGT}. It is seen that the current precision on $b_{WM}$ is sufficient to reach the level of 0.1\% on the Fierz term.

The impact of the induced tensor needs also to be considered case by case. In isospin triplets for which the $\beta^-$ and the $\beta^+$ branches can separately be studied, this form factor can be extracted experimentally, as  performed in the mass $A=20$ system \cite{Min11}. Such a comparison is impossible in mass $A=6$ because the $^6$Be ground state is unbound. It is nevertheless still possible to obtain an experimental limit on this form factor from its contribution to other observables. The actual determination of a value for the induced tensor currently relies exclusively on theory. Because of the simple properties of the allowed $^6$He decay, the induced tensor form factor is expected to be strongly suppressed \cite{Cal75}, which has been recently confirmed using state of the art ab-initio methods\cite{Navrat16}.

For mirror transitions between $T=1/2$ doublets, the weak magnetism form factor is predicted from CVC by the difference between the magnetic moments of the initial and final states, whereas the induced tensor is strongly suppressed by isospin symmetry. In nuclei with $T=\frac{3}{2}$ second rank tensor-order recoil corrections are also almost entirely given by CVC predictions, or suppressed by isospin symmetry.

\begin{table}[!htb]
\caption{Sensitivities to hadronic form factors of the Fierz term, contributing to the $\beta$ energy spectrum in $^6$He decay.}
\begin{ruledtabular}
\begin{tabular}{lcc}
Parent nucleus &
$\Delta b_{GT}/\Delta b_{WM}$ & $\Delta b_{GT}/\Delta d^{I}$ \\ 
\hline
$^6$He & $5.7\times10^{-4}$ & $1.9\times10^{-5}$
\end{tabular}
\label{tab:Deltas_bGT}
\end{ruledtabular}
\end{table}

\section{Nuclear Theory in the Next Decade}
\label{sec: nuclear theory next decade}
The previous section has laid out the main challenges of nuclear theory, which include three main fronts: (i) nuclear structure effects in the calculation of radiative corrections, particularly the $\gamma$-W box; (ii) nuclear structure corrections to the interaction of the electro-weak probes (the $\beta$ particle and the neutrino or anti-neutrino) with the nucleus, beyond the leading order approximation of the probes interacting with a single nucleon in the nucleus; and (iii) a lattice-QCD assessment of nucleon charges, 
essential to connect nuclear observables to quark-level couplings. 
In particular, the uncertainties in 
$g_A$, $g_{S}$, and $g_T$, limit the sensitivity 
to $\epsilon_R$, $\epsilon_S$, and $\epsilon_T$, respectively.

We first elaborate on the nuclear structure effects in (i) and (ii), and emphasize the current and near-future abilities of nuclear structure calculations.

The feasibility to overcome these challenges relies on much progress made in the last few years in the field of ab-initio nuclear calculations, allowing precision evaluation of nuclear wave functions and matrix elements, from light nuclei to nuclei in the mass range of $A=100$. These calculations use a nuclear potential, which is systematically built using effective field theory (EFT) of QCD at low energies (for a review see \cite{Hergert:2016nr}). The use of EFT expansion enables an assessment of the inherent accuracy of the calculation, a property which is very attractive in the context of searching for beyond the standard model signatures. Uncertainty quantification for nuclear calculations has seen a huge progress in recent years \cite{Furnstahl_2015,Wesolowski_2019}. In addition, this EFT is used to construct consistently the interaction of a weak probe with a nucleus.

A recent study can be used as an example of the current state-of-the-art in the computation of $\beta$-decays from up to $^{100}$Sn. This work combines the aforementioned effective field theories of the strong and weak forces, with powerful quantum many-body techniques \cite{Gysbers:2019}. The results indicate that accuracy of a few percent, with high precision wave functions, can be achieved in the Gamow-Teller strengths, at nuclear masses relevant to the $\beta$-decay searches mentioned here. This work shows that there is no need to suppress the axial constant artificially in order to reproduce experimental Gamow-Teller strengths, and thus gives a solution to the long-standing problem of ``$g_A$ quenching'' in nuclear systems. The same study presents that this feasibility can be achieved with other fully ab-initio methods such as the No-Core Shell Model \cite{BARRETT2013131}, and the valence-space in-medium similarity renormalization group (VS-IMSRG) method \cite{PhysRevLett.118.032502}, for light and medium mass nuclei.
Moreover, similar findings have been obtained  using Quantum Monte Carlo methods for light nuclei ($A=6-10$) in Ref.~\cite{Pastore:2017uwc}, as well as expansions in hyper-spherical harmonics \cite{PhysRevC.79.065501}.

The success of consistent currents and forces in describing transition strengths has been demonstrated, albeit only at masses $A<15$, in the electromagnetic sector as well (see, e.g., a recent review \cite{bacca:2014es} and references therein), where similar accuracy has been found for $M_1$ transitions and nuclear magnetic moments. 

In particular, regarding the nuclear structure characterization needed for a correct interpretation of $\beta$-decay experiments: 

{\it Nuclear Structure Effects in the $\gamma$-W box}-. The common analysis of the $\gamma$-W box misses some of the strength, related to the excitations of the nucleus in the loop. This resembles the nuclear-structure corrections in muonic atoms \cite{ji:2018jp}, where a similar problem was solved by representing it as a sum rule, and using the Lanczos approach for its calculation. Although this was accomplished to date only for very light nuclei, sum rules calculations were generalized to coupled cluster methods and applied for heavier nuclei, e.g., $^{48}$Ca \cite{Hagen:2015ri}. The successful use of such methods indicates the feasibility of making progress in the calculation of nuclear structure effects for the $\gamma$-W-box radiative correction to $\beta$-decay observables.

{\it Nuclear Structure Effects in the Interaction of a Weak probe and a Nucleus}-.  It is well known that the weak interaction with a nucleus is not just a coherent sum over its nucleonic constituents. In fact, already the ''independent particle approach'' \cite{Holstein:1974gl}, introduces recoil corrections $\epsilon_r=\frac{q}{M_N}$, due to the momentum $q$ transferred to the nucleon, of mass $M_N$, in the interaction. The nucleonic kinetic energy $P$, when bound to a nucleus, introduces additional, relativistic corrections $\epsilon_{NR}=\frac{P}{M_N}$. In addition, the finite size of the nucleus introduces the multipole expansion parameter, $\epsilon_{qR}=qR$. These parameters are taken into account in some of the studies of $\beta$-decays, e.g., Ref.~\cite{Hayen:2018sd}. 
For a typical $\beta$-decay kinematics with endpoint of $Q_{max}=2$ MeV, and a medium mass nucleus (relevant to approximate $P$), one sees that all these are in fact small parameters: $\epsilon_{NR}\approx\frac{1}{5}$, $\epsilon_r\approx0.01\cdot A^{1/3}$, $\epsilon_{qR}\approx 0.002$. 

Finally, the most interesting factor is related to the fact that the strong interaction introduces many body correlations, beyond the single particle, impulse approximation. At low energies, this is modelled through an effective field theory expansion of the strong interaction at nuclear energies, with expansion parameter of $\epsilon_{EFT}\approx \frac{1}{5} \to \frac{1}{3}$. The interaction of a weak-probe of polar-vector symmetry (e.g., weak magnetism) is corrected by $\epsilon_{EFT}$, and that of a weak-probe of axial-vector symmetry (e.g., Gamow-Teller) is suppressed by more than $\epsilon_{EFT}^{2}$. The aforementioned study \cite{Gysbers:2019} uses the latter model and reaches a few-percent of accuracy. 

The fact that the nuclear model allows a few-percent accuracy on specific transitions, and that the kinematical suppression factor is small, means that future designated calculations of nuclear structure corrections to different $\beta$-decay observables can reach the needed few times $10^{-4}$ accuracy. 

We notice that identifying beyond the Standard Model effects can be done just by comparing theoretical calculations with the experimental value. However, in order to relate such differences with high energy physics, one needs to relate the different nucleonic form factors with the fundamental quark charges. For example, right-handed currents would result in a deviation of the observed axial coupling $\tilde{g}_A$ from that predicted by lattice QCD calculations. The comparison is presently limited by the precision of lattice QCD calculations, now reaching into the 1\% level~\cite{gA-nature:2018}. 
BSM signatures of non-$V-A$ structures, e.g., tensor, will be proportional to the nucleonic charge corresponding to these structures. Recent QCD calculation, Ref.~\cite{Collaboration:2016mz}, has demonstrated that such nucleonic couplings have natural values, with $\approx 5-10\%$ precision. Further improvements would put 
stronger constraints on the BSM couplings, for a given (positive or null) experimental result.

The nuclear structure parts of this work, which was described above, seems feasible, by similarity to other problems and to calculations in the literature. However, there is still much work needed to reach successful predictions for either the $\gamma-W$ box or the probe-nucleus interaction nuclear structure, a.k.a recoil, effects. In both cases, the needed work can be divided into different steps. The first step is to construct the formalism, that is, identify the operators which need to be calculated, and how these can be combined into the calculation of an observable. This effort, is still not resolved in the case of the $\gamma-W$ box, while almost fully solved for the nuclear-structure recoil type corrections. 
The following step is then to calculate the wave functions. The complexity of this step, in the case of precision beta decay studies is due to the fact that these are not very light nuclei, and thus one needs to use large-scale ab-initio calculations. In the nuclear structure recoil effects this is a matter of enough computing power and person power to operate them. For the calculation of the radiative corrections, this will probably include sum rules, which complicates significantly the calculations, though still feasible \cite{Hagen:2015ri}.  

Clearly, there is a strong connection between experiment and theory in the $\beta$-decay precision front. Moreover, we expect that as new calculations accumulate, the theoretical part can point towards new directions, that would constrain Beyond the Standard Model physics in other $\beta$-decaying nuclei. For example, recent study has highlighted the relevance of forbidden decays in this realm \cite{Glick-Magid:2017vr}.
\section{Conclusions}
\label{sec: conclusions}
Improvements in nuclear beta-decay experimental techniques allow for sensitivities to new physics reaching beyond the 10 TeV scale. This is an interesting complement to the LHC program.
New developments for beta-spectra measurements and some correlations in nuclei could yield high sensitivity to new physics.
The experiments, however, cannot reach ultimate sensitivities without help from nuclear theory. Theory is needed to precisely extract $V_{ud}$ and test the unitarity of the CKM matrix, and to predict the Standard Model contributions to experiments searching for non-$V-A$ currents.

The test of the unitarity of the CKM matrix has shown recent interesting developments. Using the most recent evaluations of corrections for extraction of $V_{ud}$ yields a $> 4\sigma$ discrepancy with unitarity. The data is based on performing isospin-symmetry breaking and radiative corrections to nuclei from $^{10}{\rm C}$ to $^{74}{\rm Rb}$, and there is some concern on the accuracy of the approximations that have been used so far. 

Fortunately, recent progress in ab-initio--EFT nuclear theory indicate the calculations, albeit non trivial, should be doable with the required precision. Support for theory is thus important for the field to achieve its potential for discovery.

Overall, we see the next decade as a real opportunity to 
realize a vibrant program in precision beta decays 
with broadband
sensitivity to new physics for the full set of V, A, S and T 
couplings, probing energy scales comparable or higher than those accessible by the LHC. The envisioned research program also has the necessary ingredients to either resolve or sharpen the current CKM unitarity and neutron lifetime issues.

\begin{acknowledgments}
We gratefully acknowledge the support of the Amherst Center for Fundamental Interactions at Amherst, Massachusetts, which supported a workshop Nov. 1-3, 2018, and the European Centre for Theoretical Studies in Nuclear Physics and Related Areas, which supported a workshop at Trento, April 8-12, 2019. 

We thank John Behr, Misha Gorshtein, Geoff Greene, Daniel Phillips, Chien-Yeah Seng, Fred Wietfeldt for helpful comments.

This work is supported in part by the US Department of Energy, Office of Nuclear Physics (DE-AC02-06CH11357,  DE-FG02-97ER41020, DE-FG02-97ER41042, DE-AC52-06NA25396) and the National Science Foundation (PHY-1565546, PHY-1615153) and by the Israel Science Foundation (1446/16).

\end{acknowledgments}

\bibliography{betadecay}

\end{document}